\documentclass[12pt]{article}

\usepackage{latexsym}
\usepackage[tbtags]{amsmath}

\newcommand{\black}[1]{\boldsymbol{#1}}

\begin{document}

\title{\Large {\bf Energy-momentum non-conservation on noncommutative
        spacetime and the existence of infinite spacetime dimension  } }

\author{{\large Zheng Ze Ma}
  \\  \\ {\normalsize {\sl Department of Physics, Southeast University,
         Nanjing, 210096, P. R. China } }
  \\     {\normalsize {\sl E-mail:} z.z.ma@seu.edu.cn }}

\date{}

\maketitle

\vskip 20pt

\baselineskip 14 pt

\noindent A{\scriptsize BSTRACT}:  From the constructions of the
quantum spacetime, a four dimensional quantized spacetime can be
embedded in a five dimensional continuous spacetime. Thus to observe
from the five dimensional continuous spacetime where the four
dimensional quantized spacetime is embedded, there exist the
energy-momentum flows between the five dimensional continuous
spacetime and the four dimensional quantized spacetime. This makes
the energy-momentum not locally conserved generally on the four
dimensional quantized spacetime. We propose that energy-momentum
tensors of noncommutative field theories constructed from the
Noether approach are just the correct forms for the energy-momentum
tensors of noncommutative field theories. The non-vanishing of the
total divergences of the energy-momentum tensors of noncommutative
field theories just reflect that energy-momentum are not locally
conserved on noncommutative spacetime. At the same time, from the
constructions of the quantum spacetime, we propose that the total
spacetime dimension of the quantum spacetime is infinite.

\vskip 0.8cm

\noindent K{\scriptsize EYWORDS}: Non-Commutative Geometry,
Space-Time Symmetries

\newpage

\baselineskip 14 pt

\section{Introduction}

\indent

  People are now still not knowing that whether spacetime is
surely discrete and quantized under a very small microscopic scale.
However, from certain fundamental principles of physics, it is
possible that spacetime may have quantized and noncommutative
structures under a very small microscopic scale \cite{1,2}. This
conclusion can also be derived from superstring theories \cite{3,4}.
Field theories on noncommutative spacetime are thus called
noncommutative field theories. A lot of research works have been
carried out in this area \cite{5,6}.

  Noncommutative field theories have many different properties
from field theories on commutative spacetime. At the same time, some
questions in noncommutative field theories make people feel
delusive. One problem among them is about the energy-momentum
tensor. It is well know that there are some difficulties to
construct a locally conserved energy-momentum tensor for a
noncommutative field theory. This problem was first discovered in
noncommutative $\varphi^{4}$ scalar field theory \cite{7,8}. The
energy-momentum tensor derived from the Noether approach does not
satisfy the conservation equation. Soon after that, the
energy-momentum tensors of noncommutative electromagnetic field
theory and noncommutative gauge field theory were investigated by
several different authors \cite{9,10,11}. The same problem has been
discovered. The energy-momentum tensors of noncommutative
electromagnetic field and Yang-Mills field constructed from the
usual approach do not satisfy the conservation equation.

  Some modified forms for the energy-momentum tensors of
noncommutative field theories have been constructed. In \cite{9}, a
locally conserved energy-momentum tensor for noncommutative
$\varphi^{4}$ scalar field theory was constructed. However, it is
not symmetric. In \cite{11}, a locally conserved energy-momentum
tensor for noncommutative electromagnetic field was constructed.
However, it is not symmetric and traceless. A locally conserved
non-symmetric energy-momentum tensor has also been constructed in
\cite{12} for the renormalizable noncommutative Grosse-Wulkenhaar
scalar field theory \cite{13}. On the other hand, with the Wilson
functional integral \cite{14,15}, a locally conserved
energy-momentum tensor for noncommutative gauge field theory has
been constructed in \cite{9}. However, it is not symmetric either.
Similarly, an energy-momentum tensor including integrals for
noncommutative $\varphi^{4}$ scalar field theory was constructed in
\cite{16}, although it is symmetric and conserved.

  In this paper, we propose a different point of view for the
energy-momentum tensor problem of noncommutative field theories.
From the constructions of the quantum spacetime, a four dimensional
quantized spacetime can be embedded in a five dimensional continuous
spacetime. Thus to observe from the five dimensional continuous
spacetime where the four dimensional quantized spacetime is
embedded, there exist the energy-momentum flows between the five
dimensional continuous spacetime and the four dimensional quantized
spacetime. This makes the energy-momentum not locally conserved
generally on the four dimensional quantized spacetime. We propose
that energy-momentum tensors of noncommutative field theories
constructed from the Noether approach are just the correct forms for
the energy-momentum tensors of noncommutative field theories. The
non-vanishing of the total divergences of the energy-momentum
tensors of noncommutative field theories just reflect that
energy-momentum are not locally conserved on noncommutative
spacetime. At the same time, from the constructions of the quantum
spacetime, we propose that the total spacetime dimension of the
quantum spacetime is infinite.

  The content of this paper is organized as follows. In
section 2, we analyze the constructions of quantum spacetime and
noncommutative field theories. In section 3, from the constructions
of quantum spacetime, we propose that the four dimensional
noncommutative spacetime can be embedded in a five dimensional
commutative spacetime. Energy-momentum tensors of noncommutative
field theories can be obtained from a map relation between field
theories on the five dimensional commutative spacetime and field
theories on the four dimensional noncommutative spacetime. In
section 4, we propose that there exist the energy-momentum flows
between the five dimensional continuous spacetime and the four
dimensional quantized spacetime. Thus to observe from the four
dimensional quantized spacetime, energy-momentum are not locally
conserved. We give a derivation for the energy-momentum tensors of
noncommutative field theories from the Noether approach parallel to
the energy-momentum tensors of corresponding field theories on
commutative spacetime. In section 5, from the constructions of the
quantum spacetime, we propose that the total spacetime dimension of
the quantum spacetime is infinite. In section 6, we give some
further discussions.

\section{Quantum spacetime and noncommutative field theory}

\indent

  A concrete model of the quantized spacetime was first constructed
by Snyder in \cite{1}. In \cite{1}, Snyder decomposes the Lorentz
rotating generators of the five dimensional Minkowski spacetime into
two sets:
$$
  {\bf x}^{\mu}=ia\left(y^{4}\frac{\partial}{\partial y_{\mu}}-
         y^{\mu}\frac{\partial}{\partial y_{4}}\right),
  \eqno{(1)}  $$
$$
  {\bf M}^{\mu\nu}=-i\left(y^{\mu}\frac{\partial}{\partial y_{\nu}}
         -y^{\nu}\frac{\partial}{\partial y_{\mu}}\right),
  \eqno{(2)}  $$
where $\mu,\nu=0,1,2,3$. As the generators of the $SO(4,1)$ Lorentz
group, they satisfy the commutation relations
$$
  [{\bf x}^{\mu},{\bf x}^{\nu}]=ia^{2}{\bf M}^{\mu\nu},
  \eqno{(3)}  $$
$$
  [{\bf M}^{\mu\nu},{\bf x}^{\lambda}]=i({\bf x}^{\mu}
         \eta^{\nu\lambda}-{\bf x}^{\nu}\eta^{\mu\lambda}) ~,
  \eqno{(4)}  $$
$$
  [{\bf M}^{\mu\nu},{\bf M}^{\alpha\beta}]=i({\bf M}^{\mu\beta}
         \eta^{\nu\alpha}-{\bf M}^{\mu\alpha}\eta^{\nu\beta}+
         {\bf M}^{\nu\alpha}\eta^{\mu\beta}-
         {\bf M}^{\nu\beta}\eta^{\mu\alpha}) ~.
  \eqno{(5)}  $$
One can see that (5) is just the commutation relations for the
generators of the $SO(3,1)$ Lorentz group, i.e., the Lorentz group
of the four dimensional Minkowski spacetime. Equation (3) shows that
the coordinates of the four dimensional Minkowski spacetime are
quantized. They have the discrete spectra. In fact, because ${\bf
M}^{\mu\nu}$ are anti-symmetric for the indexes ${\mu}$ and ${\nu}$,
the time coordinate still has the continuous spectrum. Thus Snyder
has constructed a quantized spacetime in compatible with the Lorentz
invariance with its space coordinates possessing the discrete
spectra.

  In \cite{2}, to combine the principles of general relativity and
quantum mechanics, Doplicher {\sl et al.} defined a new algebra for
the quantized spacetime which is given by
$$
  [{\bf x}^{\mu},{\bf x}^{\nu}]=i\black{\theta}^{\mu\nu},
  \eqno{(6)}  $$
$$
  [\black{\theta}^{\mu\nu},{\bf x}^{\lambda}]=0 ~,
  \eqno{(7)}  $$
$$
  [\black{\theta}^{\mu\nu},\black{\theta}^{\alpha\beta}]=0 ~.
  \eqno{(8)}  $$
Similar relations can also be derived from superstring theories
\cite{3,4}. Carlson {\sl et al.} pointed out that the DFR algebra
can be derived from the Snyder algebra through a limit process
\cite{17}. To define
$$
  {\bf M}^{\mu\nu}=\frac{1}{b}\black{\theta}^{\mu\nu},
  \eqno{(9)}  $$
then to take the limit
$$
  a\rightarrow 0 ~, ~~~~~~ b\rightarrow 0 ~,
  \eqno{(10)}  $$
together with the ratio of $a^{2}$ and $b$ being fixed
$$
  \frac{a^{2}}{b}\rightarrow 1 ~,
  \eqno{(11)}  $$
the Snyder algebra can be contracted to the DFR algebra. Such a
contraction implies that there is a relation between the DFR's
quantum spacetime and Snyder's quantum spacetime; although from the
group theory language, such a contraction is not standard, as
pointed out in \cite{18}. However, to explore the relation between
the DFR's quantum spacetime and Snyder's quantum spacetime in detail
is not the purpose of this paper.

  For noncommutative field theories, they are formulated based on
the quantization of spacetime coordinates of (6)--(8). Through the
Weyl-Moyal correspondence, every field $\phi(\bf x)$ defined on the
noncommutative spacetime is mapped to its Weyl symbol $\phi(x)$
defined on the corresponding commutative spacetime. Meanwhile, the
products of field functions are replaced by the Moyal
$\star$-products of their Weyl symbols
$$
  \phi({\bf x})\psi({\bf x})\rightarrow \phi(x)\star\psi(x) ~,
  \eqno{(12)}  $$
with the Moyal $\star$-product defined as
\setcounter{equation}{12}
\begin{eqnarray}
\!\!\!\!\!\!\!\!\!\!\!\!
  \phi(x)\star\psi(x) & = & e^{\frac{i}{2}\theta^{\mu\nu}
        \frac{\partial}{\partial
        \alpha^\mu}\frac{\partial}{\partial\beta^{\nu}}}
        \phi(x+\alpha)
        \psi(x+\beta)\vert_{\alpha=\beta=0} \nonumber     \\
        & = & \phi(x)\psi(x)+\sum\limits^{\infty}_{n=1}
        \left(\frac{i}{2}\right)^{n}
        \frac{1}{n!}\theta^{\mu_{1}\nu_{1}}
        \cdots\theta^{\mu_{n}\nu_{n}}
        \partial_{\mu_{1}}\cdots\partial_{\mu_{n}}\phi(x)
        \partial_{\nu_{1}}\cdots\partial_{\nu_{n}}\psi(x) ~;
\end{eqnarray}
and the commutators of coordinate operators of (6) are equivalently
replaced by the Moyal $\star$-product commutators of the
noncommutative coordinates
$$
  [x^{\mu},x^{\nu}]_{\star}=i\theta^{\mu\nu},
  \eqno{(14)}  $$
where $\theta^{\mu\nu}$ are now the noncommutative parameters.

  From the above mentioned Weyl-Moyal correspondence, we acquaint
that the Lagrangians of field theories on noncommutative spacetime
can be formulated through replacing the products by the Moyal
$\star$-products in the Lagrangians of field theories on commutative
spacetime. For example for the Lagrangian of $\varphi^{4}$ scalar
field theory on noncommutative spacetime we have
$$
  {\cal L}=\frac{1}{2}\partial^{\mu}\varphi(x)\star
           \partial_{\mu}\varphi(x)-\frac{1}{2}m^{2}\varphi(x)
           \star\varphi(x)-\frac{1}{4!}\lambda\varphi(x)\star
           \varphi(x)\star\varphi(x)\star\varphi(x) ~.
  \eqno{(15)}  $$
Correspondingly, for electromagnetic field on noncommutative
spacetime, its Lagrangian is given by
$$
  {\cal L}=-\frac{1}{4}F^{\mu\nu}(x)\star F_{\mu\nu}(x) ~,
  \eqno{(16)}  $$
where
$$
  F_{\mu\nu}=\partial_{\mu}A_{\nu}(x)-\partial_{\nu}A_{\mu}(x)-
         i[A_{\mu}(x)\star A_{\nu}(x)-A_{\nu}(x)\star A_{\mu}(x)] ~.
  \eqno{(17)}  $$
The last term in the definition of $F_{\mu\nu}$ is in order to make
the noncommutative $U(1)$ gauge field theory be compatible with the
noncommutative $U(N)$ gauge field theories. Because in (15)--(17),
the coordinates $x^{\mu}$ satisfy the commutation relations (14),
these noncommutative field theories are defined on the four
dimensional quantized spacetime ${\bf x}^{\mu}$.

\section{Energy-momentum tensor on noncommutative spacetime
         through the map relation}

\indent

  On the other hand, from the operator construction of (1)--(5),
there exists a five dimensional commutative spacetime $y^{\mu}$,
$\mu=0,1,\ldots,4$. From the above constructions of the four
dimensional quantized spacetime ${\bf x}^{\mu}$, its points take the
eigenvalues of the operators (1). Thus its point set can be regarded
as a subset of the five dimensional continuous spacetime $y^{\mu}$.
This means that the four dimensional quantized spacetime ${\bf
x}^{\mu}$ can be embedded in the five dimensional continuous
spacetime $y^{\mu}$ through a certain manner. Its point set composes
a four dimensional hyper-surface of the five dimensional continuous
spacetime. Or we can regard that the four dimensional quantized
spacetime ${\bf x}^{\mu}$ is a subspace of the five dimensional
commutative spacetime $y^{\mu}$. Therefore there exist a map
relation between the five dimensional commutative spacetime
$y^{\mu}$ and the four dimensional quantized spacetime ${\bf
x}^{\mu}$.

  In the five dimensional commutative spacetime $y^{\mu}$, there exist
the commutative field theories. For example for the $\varphi^{4}$
scalar field theory, its Lagrangian is given by
$$
  {\cal L}=\frac{1}{2}\partial^{\mu}\varphi(y)
           \partial_{\mu}\varphi(y)-\frac{1}{2}m^{2}\varphi^{2}(y)
           -\frac{1}{4!}\lambda\varphi^{4}(y) ~.
  \eqno{(18)}  $$
For the electromagnetic field theory, its Lagrangian is given by
$$
  {\cal L}=-\frac{1}{4}F^{\mu\nu}(y)F_{\mu\nu}(y) ~,
  \eqno{(19)}  $$
where
$$
  F_{\mu\nu}=\partial_{\mu}A_{\nu}(y)-\partial_{\nu}A_{\mu}(y) ~.
  \eqno{(20)}  $$
These field theories are defined in the five dimensional commutative
spacetime $y^{\mu}$. From the five dimensional commutative spacetime
field theories to the four dimensional noncommutative spacetime
field theories, one only need to replace the product by the Moyal
$\star$-product in the Lagrangians. Thus there is also a map
relation between the five dimensional commutative field theories and
the four dimensional noncommutative field theories. However, we
conjecture that such a map relation not only exists in the
Lagrangians between these two kinds of field theories, but also
exists in some other expressions between these two kinds of field
theories. In fact, such a map relation also exists in the field
equations. Thus, it is reasonable to conjecture that such a map
relation also exists in the expressions of the energy-momentum
tensors of these two kinds of field theories.

  For field theories on the five dimensional commutative spacetime
$y^{\mu}$, their energy-momentum tensors can be obtained through the
Noether approach. For the $\varphi^{4}$ scalar field theory, we have
$$
  {\cal T}_{\mu\nu}=\partial_{\mu}\varphi(y)
        \partial_{\nu}\varphi(y)-g_{\mu\nu}{\cal L}(y) ~.
  \eqno{(21)}  $$
It is symmetric and locally conserved. For the electromagnetic
field, through the canonical approach and the Belinfante procedure,
its energy-momentum tensor can be constructed as
$$
  {\cal T}_{\mu\nu}=\frac{1}{4}g_{\mu\nu}F^{\alpha\beta}(y)
         F_{\alpha\beta}(y)-F_{\mu\lambda}(y)F_{\nu}^{\lambda}(y) ~.
  \eqno{(22)}  $$
It is symmetric, traceless, locally conserved, and also gauge
invariant. From the map relation between the five dimensional
commutative field theories and four dimensional noncommutative field
theories, we can write down the energy-momentum tensors of the
corresponding field theories on four dimensional noncommutative
spacetime. For $\varphi^{4}$ scalar field theory we have
$$
  {\cal T}_{\mu\nu}=\frac{1}{2}(\partial_{\mu}\varphi(x)\star
          \partial_{\nu}\varphi(x)+\partial_{\nu}\varphi(x)\star
          \partial_{\mu}\varphi(x))-g_{\mu\nu}{\cal L}(x) ~.
  \eqno{(23)}  $$
In (23), an additional symmetrizing procedure has been made in order
to make the expression symmetric for the reason that Moyal
$\star$-product is not symmetric between two different functions.
For the electromagnetic field, we have
$$
  {\cal T}_{\mu\nu}=\frac{1}{4}g_{\mu\nu}F^{\alpha\beta}(x)\star
         F_{\alpha\beta}(x)-\frac{1}{2}(F_{\mu\lambda}(x)\star
         F_{\nu}^{\lambda}(x)+F_{\nu\lambda}(x)\star
         F_{\mu}^{\lambda}(x)) ~.
  \eqno{(24)}  $$
The same reason as above, an additional symmetrizing procedure has
been made. Such an energy-momentum tensor is symmetric and traceless
\cite{10}. Under the gauge transformation $\delta
A_{\mu}=D_{\mu}\Omega$, it transforms covariantly
$$
  {\cal T}_{\mu\nu}\rightarrow\Omega\star{\cal T}_{\mu\nu}\star
          \Omega^{-1} ~.
  \eqno{(25)}  $$
The energy-momentum tensors in the form of (23) and (24) for
noncommutative $\varphi^{4}$ scalar field and noncommutative
electromagnetic field have been obtained in \cite{7,8,9,10,11} from
several different methods. Here we point out that we can obtain them
from a map relation between field theories on five dimensional
commutative spacetime and four dimensional noncommutative spacetime.
This insinuate that (23) and (24) may be the correct forms for the
energy-momentum tensors of noncommutative $\varphi^{4}$ scalar field
theory and noncommutative electromagnetic field theory.

  The spectrum of time coordinate in the quantized spacetime
model of Snyder is still continuous. A generalized quantum spacetime
algebra with the discrete spectrum for the time coordinate has been
constructed in \cite{19}. To adopt the construction of \cite{19}, a
four dimensional quantized spacetime with both its time and space
coordinates discrete can be embedded in a six dimensional continuous
spacetime. However in order to make the discussion of this paper
simple, we will not refer to the construction of \cite{19} in the
following.

\section{Energy-momentum non-conservation on noncommutative
spacetime}

\subsection{Energy-momentum flows between four dimensional quantized
spacetime and five dimensional continuous spacetime}

\indent

  For the energy-momentum tensors in the forms of (23) and (24) for
noncommutative $\varphi^{4}$ scalar field and noncommutative gauge
field, they do not satisfy the local conservation equation. Some
modification schemes have been proposed in order to obtain a locally
conserved energy-momentum tensor for a noncommutative field theory
\cite{9,11,12}. However, it seems that the properties of symmetry,
traceless, and gauge covariance usually cannot be satisfied together
for the modified forms of the energy-momentum tensors of
noncommutative field theories. In \cite{16}, a modified form for the
energy-momentum tensor of noncommutative $\varphi^{4}$ scalar field
theory has been constructed. It is symmetric and conserved. However,
its expression includes path-dependent integrals. It is not the
ideal form for the energy-momentum tensor of noncommutative
$\varphi^{4}$ scalar field theory.

  In this paper, we propose that energy-momentum are not locally
conserved for a field theory on noncommutative spacetime. We propose
that energy-momentum tensors of noncommutative field theories
constructed from the Noether approach are the correct forms for the
energy-momentum tensors of noncommutative field theories. In order
to arrive our proposal, we look at this problem from a different
angle. In section 2 we have seen that from the constructions of the
four dimensional quantized spacetime ${\bf x}^{\mu}$, its points
take the eigenvalues of the operators (1). Thus its point set can be
regarded as a subset of the five dimensional continuous spacetime
$y^{\mu}$, the four dimensional quantized spacetime ${\bf x}^{\mu}$
can be embedded in the five dimensional continuous spacetime
$y^{\mu}$ through a certain manner, its point set composes a
subspace of the five dimensional continuous spacetime $y^{\mu}$. Now
to observe from the five dimensional continuous spacetime $y^{\mu}$
where the four dimensional quantized spacetime ${\bf x}^{\mu}$ is
embedded, matter and energy-momentum are existing and moving in such
a five dimensional continuous spacetime, and the energy-momentum
have the property of local conservation. However, there exist the
exchange of matter and energy-momentum between the four dimensional
sub-spacetime ${\bf x}^{\mu}$ and the five dimensional spacetime
$y^{\mu}$. Thus to observe from the four dimensional quantized
spacetime ${\bf x}^{\mu}$, the energy-momentum are not locally
conserved necessarily. This is just like the case of a two
dimensional fluid flowing in a three dimensional space. If there are
no sources or holes in the two dimensional space, to observe from
the two dimensional space where the fluid is flowing, there exist
the continuous and conservation equations for the two dimensional
fluid. However, if there exist sources and holes in the two
dimensional space where the fluid is flowing, then to observe from
the two dimensional space, the energy-momentum of the fluid are not
locally conserved at the locations where there are sources or holes;
because at these places, there exist the energy-momentum exchange
(import or export) between the two dimensional space and three
dimensional space. However, to observe from the three dimensional
space, there still exist the continuous and conservation equations
for the fluid. For the four dimensional quantized spacetime ${\bf
x}^{\mu}$ embedded in a five dimensional continuous spacetime
$y^{\mu}$, the situation is similar. Thus there may exist the
energy-momentum import from the five dimensional continuous
spacetime $y^{\mu}$ to the four dimensional quantized spacetime
${\bf x}^{\mu}$ or the energy-momentum export from the four
dimensional quantized spacetime ${\bf x}^{\mu}$ to the five
dimensional continuous spacetime $y^{\mu}$ at every spacetime point
of the four dimensional quantized spacetime.

  Therefore, from the above analysis, we can understand that
energy-momentum are not locally conserved necessarily to observe in
the four dimensional quantized spacetime. However, to observe from
the five dimensional continuous spacetime where the four dimensional
quantized spacetime is embedded, energy-momentum are still locally
conserved. Therefore we consider that the non-vanishing of the total
divergences of the energy-momentum tensors do not mean that the
energy-momentum tensors constructed from the Noether approach are
not the correct forms for the energy-momentum tensors of
noncommutative field theories; they just imply that there exist the
energy-momentum exchange or energy-momentum flows between the four
dimensional quantized spacetime and the five dimensional continuous
spacetime.

\subsection{Open system on four dimensional commutative spacetime}

\indent

  The above analysis shows that field theories on the four dimensional
quantized spacetime are not the closed systems, due to the existing
of the local energy-momentum flows between the four dimensional
quantized spacetime and the five dimensional continuous spacetime.
They are open systems in fact. This will affect the forms of field
equations. In order to elucidate the problem, we first consider an
open system on the four dimensional commutative spacetime.

  For the general case, we can write the energy-momentum tensor in
the form
$$
  {\cal T}_{\mu\nu}={\cal T}_{\nu\mu}=\left(
  \begin{array}{llll}
     w     & g_{1}  & g_{2}  & g_{3}    \\
     S_{1} & T_{11} & T_{12} & T_{13}    \\
     S_{2} & T_{21} & T_{22} & T_{23}    \\
     S_{3} & T_{31} & T_{32} & T_{33}    \\
  \end{array}  \right),
  \eqno{(26)}  $$
where $w$ is the energy density, ${\bf S}$ is the energy flux
density, ${\bf g}$ is the momentum density, and $T_{ij}$ is the
three-dimensional stress tensor. Supposing that there exist the
adscititious local sources and flows
$$
  f_{\mu}=(e,f_{1},f_{2},f_{3})
  \eqno{(27)}  $$
for the energy-momentum on the four dimensional commutative
spacetime, then the conservation and motion equations take the
following form
$$
  \partial^{\mu}{\cal T}_{\mu\nu}=f_{\nu} ~.
  \eqno{(28)}  $$
The four equations contained in (28) can be written down explicitly:
$$
  \frac{\partial w}{\partial t}+\nabla\cdot{\bf S}=e ~,
  \eqno{(29)}  $$
$$
  \frac{\partial {\bf g}}{\partial t}+
     \nabla\cdot{\bf T}=\black{f} ~,
  \eqno{(30)}  $$
where ${\bf T}=T_{ij}$ is the stress tensor. In the right hand sides
of (29) and (30), $(e,f_{1},f_{2},f_{3})$ represent the local
adscititious energy-momentum flows on the spacetime.

  For the primal system, we suppose that it is composed of the fields
$\varphi_{r}(x), r=1,2,...,s$. We suppose that the imported
energy-momentum flows $f_{\mu}(x)$ are composed of the same matter,
i.e., they are described by the same fields $\varphi_{r}(x),
r=1,2,...,s$. We do not regard the imported energy-momentum flows
$f_{\mu}(x)$ as the external sources. We consider that they are
amalgamated with the primal system together. Thus we should use the
fields $\varphi_{r}(x)$ to describe the whole open system. It should
be correct that the local field equations
$$
  \frac{\partial{\cal L}}{\partial\varphi_{r}}-
    \frac{\partial}{\partial x_{\mu}}\frac{\partial{\cal L}}
       {\partial(\partial\varphi_{r}/\partial x_{\mu})}=0 ~,
  ~~~~~~~~ r=1,2,...,s
  \eqno{(31)}  $$
are still satisfied at every spacetime point for the whole open
system. It is also reasonable to suppose that the expression for the
local Lagrangian ${\cal L}$ is not changed for such an open system,
because the field equations (31), which is just the Euler-Lagrangian
equation, can be derived from the variation principle. Thus, for
this open system, its Lagrangian
$$
  {\cal L}={\cal L}(\varphi_{r},\partial\varphi_{r}/
           \partial x_{\mu})
  \eqno{(32)}  $$
is just the same as the Lagrangian of the closed system.

  However, when the energy-momentum flows $f_{\mu}(x)$ are imported,
the system will take an integral change. Such an integral change
reflects on the integral change of the fields $\varphi_{r}(x),
r=1,2,...,s$. To view from the primal system, such an integral
change for the fields is non-local; and also it violates the
causality. Therefore the field theory for such an open system is a
kind of non-local field theory; or it contains the component of
non-local part. However, for such a non-local field system, we can
still use the same local fields $\varphi_{r}(x)$, local Lagrangian
${\cal L} (\varphi_{r},\partial\varphi_{r}/\partial x_{\mu})$, and
local field equations (31) to describe it; only that the local
conservation and motion equations for the energy-momentum tensor are
changed. They are replaced by (28) in fact. And also the non-local
contents of this system are completely contained by (28). In fact,
we can regard (28) as the constraint equations exerted on the
system. Therefore we can adopt the method of local field theory to
study such an open non-local system; but now it is a constraint
system. We point out here that field theories on noncommutative
spacetime can just be regarded as open non-local systems like this.
Therefore such a model will be helpful for us to understand the
construction of the energy-momentum tensors for field theories on
noncommutative spacetime. For such a purpose, we will first make a
derivation for (28)--(30) from the Noether approach.

  In order to derive (28)--(30) from the Noether approach, we can
suppose that under the infinitesimal displacements of the spacetime
coordinates
$$
  x^{\prime}_{\mu}=x_{\mu}+\epsilon_{\mu} ~,
  \eqno{(33)}  $$
the Lagrangian is changed as
$$
  \delta{\cal L}={\cal L}(x^{\prime})-{\cal L}(x)+
      \epsilon_{\mu}f^{\mu}=
      \epsilon_{\mu}\frac{\partial{\cal L}}{\partial x_{\mu}}+
      \epsilon_{\mu}f^{\mu} ~.
  \eqno{(34)}  $$
Here, the additional term $\epsilon_{\mu}f^{\mu}$ in (34) is
subjected to the imported energy-momentum flows. On the other hand,
from (32), because ${\cal L}$ does not depend on the coordinates
apparently, we have
$$
  \delta{\cal L}=\frac{\partial{\cal L}}
       {\partial\varphi_{r}}\delta\varphi_{r}+
       \frac{\partial{\cal L}}
       {\partial(\partial\varphi_{r}/\partial x_{\mu})}
        \delta\left(\frac{\partial\varphi_{r}}
       {\partial x_{\mu}}\right),
  ~~~~~~~~ r=1,2,...,s ~,
  \eqno{(35)}  $$
where
$$
  \delta\varphi_{r}=\varphi_{r}(x+\epsilon)-\varphi_{r}(x)=
        \epsilon_{\mu}\frac{\partial\varphi_{r}(x)}
        {\partial x_{\mu}} ~.
  \eqno{(36)}  $$
From (34), (35), and (31), we have
$$
  \epsilon_{\mu}\frac{\partial{\cal L}}{\partial x_{\mu}}+
      \epsilon_{\mu}f^{\mu}=\frac{\partial}{\partial x_{\mu}}
      \left(\frac{\partial{\cal L}}
      {\partial(\partial\varphi_{r}/\partial x_{\mu})}
       \epsilon_{\nu}\frac{\partial\varphi_{r}}
       {\partial x_{\nu}}\right).
  \eqno{(37)}  $$
Because (37) is satisfied for an arbitrary $\epsilon_{\mu}$, we
obtain
$$
  \frac{\partial}{\partial x_{\mu}}\left(
       -g_{\mu\nu}{\cal L}+\frac{\partial{\cal L}}
       {\partial(\partial\varphi_{r}/\partial x_{\mu})}
       \frac{\partial\varphi_{r}}{\partial x_{\nu}}\right)
       =g_{\mu\nu}f^{\mu} ~.
  \eqno{(38)}  $$
Then to write (38) in the form of (28), we obtain the
energy-momentum tensor of the system
$$
  {\cal T}_{\mu\nu}=-g_{\mu\nu}{\cal L}+
       \frac{\partial{\cal L}}
       {\partial(\partial\varphi_{r}/\partial x_{\mu})}
       \frac{\partial\varphi_{r}}{\partial x_{\nu}} ~.
  \eqno{(39)}  $$
This means that for such an open system, its energy-momentum tensor
can be expressed as the same form as that of the closed system. In
fact, it is correct.

\subsection{Energy-momentum tensor on noncommutative spacetime
from the Noether approach}

\indent

  In this subsection, we give a derivation for the energy-momentum
tensors of noncommutative field theories from the Noether approach.
We consider the noncommutative field theories that only contain the
first order derivatives in their Lagrangians written in the form of
the Moyal $\star$-product. Thus their Lagrangians can be written as
$$
  {\cal L}_{\star}={\cal L}_{\star}
    (\varphi_{r},\partial\varphi_{r}/\partial x_{\mu})
  \eqno{(40)}  $$
generally, where $r=1,2,...,s$ represent different independent field
components. Although for the Lagrangians of (40), when expanded
according to the parameters $\theta^{\mu\nu}$, they contain higher
derivative terms, we consider that on noncommutative spacetime the
Moyal $\star$-product is the fundamental product operation, we need
not to expand the Moyal $\star$-product in the Lagrangians in
principle. Thus field equations of $\varphi_{r}(x)$ can still be
cast in the form of the Euler-Lagrange equation which can be derived
from the variation principle. They are given by
$$
  \frac{\partial{\cal L}_{\star}}{\partial\varphi_{r}}-
  \frac{\partial}{\partial x_{\mu}}\frac{\partial{\cal L}_{\star}}
       {\partial(\partial\varphi_{r}/\partial x_{\mu})}=0 ~,
  ~~~~~~~~ r=1,2,...,s ~.
  \eqno{(41)}  $$
For example for noncommutative $\varphi^{4}$ scalar field theory
with the Lagrangian (15), from (41), we obtain the field equation
$$
  (\Box+m^{2})\varphi+\frac{1}{3!}\lambda\varphi\star\varphi
              \star\varphi=0 ~.
  \eqno{(42)}  $$

  As analyzed above, we consider that the four dimensional
noncommutative spacetime is embedded in a five dimensional
commutative spacetime, there exist energy-momentum flows from the
five dimensional commutative spacetime to the four dimensional
noncommutative spacetime. Therefore field theories on noncommutative
spacetime are open systems. Like that of the open systems on the
four dimensional commutative spacetime, we suppose that under the
infinitesimal displacements of the spacetime coordinates
$$
  x^{\prime}_{\mu}=x_{\mu}+\epsilon_{\mu} ~,
  \eqno{(43)}  $$
the displacement of the Lagrangian is given by
$$
  \delta{\cal L}_{\star}={\cal L}_{\star}(x^{\prime})-
      {\cal L}_{\star}(x)+\epsilon_{\mu}f^{\mu}=
      \epsilon_{\mu}\frac{\partial{\cal L}_{\star}}
      {\partial x_{\mu}}+\epsilon_{\mu}f^{\mu},
  \eqno{(44)}  $$
where the term $\epsilon_{\mu}f^{\mu}$ is due to the existence of
the energy-momentum flows from the five dimensional commutative
spacetime to the four dimensional noncommutative spacetime. On the
other hand, from (40), because ${\cal L}_{\star}$ does not rely on
the coordinates apparently, we have
$$
  \delta{\cal L}_{\star}=\frac{\partial{\cal L}_{\star}}
       {\partial\varphi_{r}}\star\delta\varphi_{r}+
       \frac{\partial{\cal L}_{\star}}
       {\partial(\partial\varphi_{r}/\partial x_{\mu})}
        \star\delta\left(\frac{\partial\varphi_{r}}
       {\partial x_{\mu}}\right),
  ~~~~~~~~ r=1,2,...,s ~,
  \eqno{(45)}  $$
where
$$
  \delta\varphi_{r}=\varphi_{r}(x+\epsilon)-\varphi_{r}(x)=
        \epsilon_{\mu}\frac{\partial\varphi_{r}(x)}
        {\partial x_{\mu}} ~.
  \eqno{(46)}  $$
From (44), (45), and (41), we have
$$
  \epsilon_{\mu}\frac{\partial{\cal L}_{\star}}{\partial x_{\mu}}
      +\epsilon_{\mu}f^{\mu}=\frac{\partial}{\partial x_{\mu}}
      \left(\frac{\partial{\cal L}_{\star}}
      {\partial(\partial\varphi_{r}/\partial x_{\mu})}\star
       \epsilon_{\nu}\frac{\partial\varphi_{r}}
       {\partial x_{\nu}}\right).
  \eqno{(47)}  $$
Because (47) is satisfied for an arbitrary $\epsilon_{\mu}$, we
obtain
$$
  \frac{\partial}{\partial x_{\mu}}\left(
       -g_{\mu\nu}{\cal L}_{\star}+
       \frac{\partial{\cal L}_{\star}}
       {\partial(\partial\varphi_{r}/\partial x_{\mu})}\star
       \frac{\partial\varphi_{r}}{\partial x_{\nu}}\right)=
       g_{\mu\nu}f^{\mu} ~.
  \eqno{(48)}  $$
We can write (48) in the form
$$
  \partial^{\mu}{\cal T}_{\mu\nu}=f_{\nu} ~,
  \eqno{(49)}  $$
where
$$
  {\cal T}_{\mu\nu}=-g_{\mu\nu}{\cal L}_{\star}+
       \frac{\partial{\cal L}_{\star}}
       {\partial(\partial\varphi_{r}/\partial x_{\mu})}\star
       \frac{\partial\varphi_{r}}{\partial x_{\nu}} ~.
  \eqno{(50)}  $$
Like that of the open system on the four dimensional commutative
spacetime, we can explain (50) as the energy-momentum tensor of a
field theory on noncommutative spacetime. The non-vanishing of the
total divergence of the energy-momentum tensor is due to the
existence of the energy-momentum flows from the five dimensional
commutative spacetime to the four dimensional noncommutative
spacetime.

  To notice that the Moyal $\star$-product is not invariant generally
under the commutation of the order of two functions, we need to
write (50) in a symmetrized form with respect to the Moyal
$\star$-product. Therefore we have
$$
  {\cal T}_{\mu\nu}=-g_{\mu\nu}{\cal L}_{\star}+\frac{1}{2}\left(
       \frac{\partial{\cal L}_{\star}}{\partial(\partial\varphi_{r}/
       \partial x_{\mu})}
       \star\frac{\partial\varphi_{r}}{\partial x_{\nu}}
       +\frac{\partial\varphi_{r}}{\partial x_{\nu}}\star
        \frac{\partial{\cal L}_{\star}}{\partial(\partial\varphi_{r}/
        \partial x_{\mu})}\right).
  \eqno{(51)}  $$
In fact (51) can be resulted through symmetrizing the Moyal
$\star$-product of any two functions properly in the above
derivation. For the Lagrangian (15) of the noncommutative
$\varphi^{4}$ scalar field theory, from (51) we obtain
$$
  {\cal T}_{\mu\nu}=\frac{1}{2}(\partial_{\mu}\varphi(x)\star
          \partial_{\nu}\varphi(x)+\partial_{\nu}\varphi(x)\star
          \partial_{\mu}\varphi(x))-g_{\mu\nu}{\cal L}(x) ~,
  \eqno{(52)}  $$
which is just given by (23). For the Lagrangian (16) of the
noncommutative electromagnetic field theory, in order to obtain a
symmetric and traceless energy-momentum tensor from the above
approach, we need to invoke the Belinfante procedure at the same
time. We omit to write down the detailed process here. At last we
obtain
$$
  {\cal T}_{\mu\nu}=\frac{1}{4}g_{\mu\nu}F^{\alpha\beta}(x)\star
         F_{\alpha\beta}(x)-\frac{1}{2}(F_{\mu\lambda}(x)\star
         F_{\nu}^{\lambda}(x)+F_{\nu\lambda}(x)\star
         F_{\mu}^{\lambda}(x)) ~,
  \eqno{(53)}  $$
which is the same form of (24). Similar procedure can be applied to
the construction of the energy-momentum tensor of noncommutative
$U(N)$ gauge field theory. The final expression for the
energy-momentum tensor of noncommutative $U(N)$ gauge field theory
is just given by (53). The expressions of (52) and (53) have been
obtained previously in the literature from several different methods
\cite{7,8,9,10,11}. As pointed out in \cite{10}, the energy-momentum
tensor of (53) is symmetric, traceless, and gauge covariance.

\subsection{Energy-momentum non-conservation on noncommutative
spacetime}

\indent

  In fact, we have not derived the energy-momentum tensors of
noncommutative field theories from the above approach really.
Because we consider that the four dimensional quantized spacetime is
embedded in a five dimensional continuous spacetime, there exist
energy-momentum flows from the five dimensional continuous spacetime
to the four dimensional quantized spacetime, energy-momentum are not
locally conserved on the four dimensional quantized spacetime. In
(49), we explain $f_{\mu}(x)$ as the energy-momentum flows from the
five dimensional continuous spacetime to the four dimensional
quantized spacetime. But here, $f_{\mu}(x)$ themselves are
determined by the total divergence of the energy-momentum tensor
constructed above. For such a case, $f_{\mu}(x)$ can be expressed as
functions of $(\varphi_{r},\partial\varphi_{r}/\partial x_{\mu})$,
i.e.,
$$
  f_{\mu}=f_{\mu}(\varphi_{r},\partial\varphi_{r}/
    \partial x_{\mu}) ~.
  \eqno{(54)}  $$
We explain (50) or (51) as the energy-momentum tensors of field
theories on noncommutative spacetime. In addition, field theories on
noncommutative spacetime are open systems, they have the property of
non-locality. For closed systems on commutative spacetime,
Poincar{\' e} translation invariance results the existence of
locally conserved energy-momentum tensors. Thus (49) also implies
that Poincar{\' e} translation invariance is violated for field
theories on noncommutative spacetime.

  There are some other reasons supporting us to explain (51)--(53) as
the correct forms of the energy-momentum tensors of noncommutative
field theories. First as pointed out in section 3, the
energy-momentum tensors expressed by (51)--(53) can be obtained
directly from a map relation between field theories on five
dimensional commutative spacetime and four dimensional
noncommutative spacetime. The second reason is that the
energy-momentum tensor for a noncommutative field theory constructed
from the above approach is symmetric; for gauge field theory, to
combine the Belinfante mechanism together, it is symmetric,
traceless, and gauge covariance. The third reason is that when we
take the limit $\theta^{\mu\nu}\rightarrow 0$ in (51)--(53), which
is equivalent to replace the Moyal $\star$-product by the ordinary
product, these expressions for the energy-momentum tensors of
noncommutative field theories come back to the energy-momentum
tensors of corresponding field theories on commutative spacetime.

  The explicit forms of $f_{\mu}(x)$ for a noncommutative field
theory can be obtained from (49) and (51). For noncommutative
$\varphi^{4}$ scalar field theory, from (52), we have \cite{7,8}
$$
  \partial^{\mu}{\cal T}_{\mu\nu}=\frac{\lambda}{4!}\left[
     \left[\varphi,\partial_{\nu}\varphi\right]_{\star},
     \varphi^{\star 2}\right]_{\star} ~.
  \eqno{(55)}  $$
For the energy-momentum tensor of noncommutative gauge field theory
given by (53), it satisfies the covariant conservation equation
\cite{9,10,11}
$$
  D_{\mu}\star{\cal T}^{\mu\nu}=\partial_{\mu}
     {\cal T}^{\mu\nu}-i(A_{\mu}\star{\cal T}^{\mu\nu}-
     {\cal T}^{\mu\nu}\star A_{\mu})=0 ~,
  \eqno{(56)}  $$
thus we have
$$
  \partial_{\mu}{\cal T}^{\mu\nu}=i(A_{\mu}\star{\cal T}^{\mu\nu}-
     {\cal T}^{\mu\nu}\star A_{\mu}) ~.
  \eqno{(57)}  $$
For the noncommutative electromagnetic field, to the first order of
$\theta^{\mu\nu}$, one obtains \cite{11}
$$
  \partial_{\mu}{\cal T}^{\mu\nu}=\partial_{\mu}\left[(\theta
           F{\cal T}^{(0)})^{\mu\nu}-\partial_{\beta}
            (\theta^{\alpha\beta}A_{\alpha}
            {\cal T}^{(0)\mu\nu})\right],
  \eqno{(58)}  $$
where ${\cal T}^{(0)\mu\nu}$ is the energy-momentum tensor of
electromagnetic field on commutative spacetime, i.e., to let
$\theta^{\mu\nu}=0$ in (53).

  To observe from the four dimensional
noncommutative spacetime only, we can also explain
$\partial^{\mu}{\cal T}_{\mu\nu}$ as the local generation of the
energy-momentum on the four dimensional noncommutative spacetime.
For the energy-momentum tensor of (51), the four-momentum over the
whole three-dimensional space is given by
$$
  P_{\mu}=\int d^{3}x{\cal T}_{0\mu} ~.
  \eqno{(59)}  $$
For noncommutative $\varphi^{4}$ scalar field theory we have
$$
  E=\int d^{3}x{\cal T}_{00}=\int d^{3}x (\pi\star\pi-{\cal L})
         =\int d^{3}x(\pi\star\stackrel{\cdot}{\varphi}-{\cal L}) ~,
  \eqno{(60)}  $$
$$
  P_{i}=\int d^{3}x{\cal T}_{0i}=\int d^{3}x\frac{1}{2}
        (\pi\star\partial_{i}\varphi+\partial_{i}\varphi\star\pi) ~,
  \eqno{(61)}  $$
where $\pi=\partial{\cal L}/
\partial\stackrel{\cdot}{\varphi}=\stackrel{\cdot}{\varphi}$ is
the conjugate momentum. The four-momentum can be written as
$P^{\mu}=(E,P_{1},P_{2},P_{3})$. Because
$$
  \int d^{3}x\partial^{\mu}{\cal T}_{\mu\nu}=\partial^{0}\int d^{3}x
            {\cal T}_{0\nu}+\int d^{3}x\partial^{i}{\cal T}_{i\nu}=
             \partial^{0}\int d^{3}x{\cal T}_{0\nu} ~,
  \eqno{(62)}  $$
we have
$$
  \partial^{0}P_{\mu}=\int d^{3}x\frac{\lambda}{4!}\left[\left[\varphi,
            \partial_{\mu}\varphi\right]_{\star},
            \varphi^{\star 2}\right]_{\star} ~.
  \eqno{(63)}  $$
We explain (63) as the generation of energy-momentum over the whole
three-dimensional space for the noncommutative $\varphi^{4}$ scalar
field theory. It is also necessary to point out that for the
space-space noncommutativity with $\theta^{0i}=0$, the right hand
side of (63) disappears due to the integral property of the Moyal
$\star$-product on the three-dimensional space \cite{7}. For such a
case, the energy-momentum defined by (59) for noncommutative
$\varphi^{4}$ scalar field theory are totally conserved. However,
the right hand side of (55) is still non-vanishing generally. Thus
the energy-momentum of the noncommutative $\varphi^{4}$ scalar field
are not locally conserved even if $\theta^{0i}=0$.

  We can give another argument for our proposal from the property
of the propagation of nonlinear waves on noncommutative spacetime.
From \cite{20,21}, we know that nonlinear perturbations and waves of
fields on noncommutative spacetime have infinite propagation speed.
From such a fact, we can understand it is possible that
energy-momentum are not locally conserved on noncommutative
spacetime. This is because usually local conservation of
energy-momentum is a property for waves of finite propagation speed.
When the propagation speed of a wave is infinite, one cannot exert
the property of local conservation of energy-momentum on such a
wave. For noncommutative $\varphi^{4}$ scalar field, we can see that
the right hand side of (55) is zero for a plane wave. Thus
energy-momentum are locally conserved for a plane wave of the
noncommutative $\varphi^{4}$ scalar field. This is in accordance
with the conclusion of \cite{21} where it is pointed out that the
infinite propagation speed is a property of nonlinear waves for
scalar field on noncommutative spacetime. For the electromagnetic
field on noncommutative spacetime, from (58) we can see that the
total divergence of the energy-momentum tensor is not zero even for
a plane wave. We can understand this from the non-Abelian structure
(cf. (17)) of the $U(1)$ gauge field on noncommutative spacetime.
Because of the existence of the non-Abelian structure, there does
not exist the exact plane wave for the electromagnetic field on
noncommutative spacetime in fact. Thus the waves of electromagnetic
field on noncommutative spacetime are always nonlinear. They have
infinite propagation speed like the nonlinear waves of scalar field
on noncommutative spacetime \cite{21}. Thus even if for a plane wave
of the $U(1)$ gauge field on noncommutative spacetime, the local
energy-momentum generation are not zero.

\section{The existence of infinite spacetime dimension}

\indent

  In the previous sections, we have proposed that the four
dimensional quantized spacetime ${\bf x}^{\mu}$ can be considered as
a subspace embedded in a five dimensional continuous spacetime
$y^{\mu}$, and there exist energy-momentum flows between the four
dimensional quantized spacetime ${\bf x}^{\mu}$ and the five
dimensional continuous spacetime $y^{\mu}$. Thus energy-momentum are
not locally conserved on the four dimensional quantized spacetime
${\bf x}^{\mu}$. However this is not the end of the problem.

  Now we consider the five dimensional continuous spacetime
$y^{\mu}$. In such a classical and continuous spacetime, there still
exist the Einstein gravitational theory and quantum theory. The
principles of general relativity and quantum mechanics still take
effect. According to the arguments of \cite{2}, such a spacetime
cannot be a classical existence in fact. It should also be quantized
under a very small microscopic scale such as the Planck scale. Thus
the DFR algebra (6)--(8) should also be exerted on such a five
dimensional continuous spacetime; and we will obtain a quantized
five dimensional spacetime ${\bf y}^{\mu}$ at last. Field theories
on such a five dimensional spacetime now also become noncommutative
field theories. Thus on such a five dimensional quantized spacetime,
there exist the five dimensional noncommutative scalar field theory
and noncommutative electromagnetic field theory. Their Lagrangians
are still given by the forms of (15) and (16); and their
energy-momentum tensors are still given by the forms of (23) and
(24). Similar as the case of the four dimensional noncommutative
spacetime, energy-momentum are not locally conserved on such a five
dimensional quantized spacetime.

  On the other hand, from the constructions of quantum
spacetime as analyzed in section 2, such a five dimensional
quantized spacetime ${\bf y}^{\mu}$ can be embedded in a six
dimensional continuous spacetime $w^{\mu}$. Such a six dimensional
continuous spacetime has the Lorentz group $SO(5,1)$. In such a six
dimensional continuous spacetime $w^{\mu}$, there exist the
commutative field theories, such as scalar field theory and
electromagnetic field theory. And there exist energy-momentum flows
between the five dimensional quantized spacetime ${\bf y}^{\mu}$ and
the six dimensional continuous spacetime $w^{\mu}$. Thus
energy-momentum are not locally conserved on the five dimensional
quantized spacetime ${\bf y}^{\mu}$.

  Such an inference can be proceeded to an arbitrary spacetime
dimension $n$. Thus we can deduce that there exists an
$n$-dimensional quantized spacetime ${\bf X}^{\mu}$, it is embedded
in an $(n+1)$-dimensional continuous spacetime $Y^{\mu}$. However,
the $(n+1)$-dimensional continuous spacetime $Y^{\mu}$ also needs to
be quantized according to the arguments of \cite{2}. Therefore at
last, $n$ will tend to the limit of $\infty$. Thus the total
spacetime dimension is infinite. And at the same time, the infinite
dimensional spacetime is quantized. Its coordinates satisfy the DFR
algebra. Field theories on such an infinite dimensional spacetime
are noncommutative field theories. There exist the energy-momentum
flows between the $\infty$-dimensional quantized spacetime
$\black{{\cal X}}^{\mu}$ and the $(\infty+1)$-dimensional continuous
spacetime ${\cal Y}^{\mu}$. However we can consider that
$\infty=\infty+1$. Therefore to observe from the whole infinite
dimensional spacetime, there does not exist the local
energy-momentum non-conservation. Thus the local conservation of
energy-momentum for field theories on noncommutative spacetime can
be realized in an $\infty$-dimensional quantum spacetime. That is to
say, if we demand that the energy-momentum should be locally
conserved for physics, then the total dimension of the spacetime
should be infinite, if the spacetime is quantized and
noncommutative.

\section{Some further discussions}

\indent

  In this paper, from the constructions of the quantum spacetime, we
propose that a four dimensional quantized spacetime can be embedded
in a five dimensional continuous spacetime. Thus to observe from the
five dimensional continuous spacetime where the four dimensional
quantized spacetime is embedded, there exist the energy-momentum
flows from the five dimensional continuous spacetime to the four
dimensional quantized spacetime, energy-momentum are not locally
conserved on the four dimensional quantized spacetime. Thus
noncommutative field systems are not closed systems; they are open
systems in fact. We propose that energy-momentum tensors of
noncommutative field theories constructed from the Noether approach
are just the correct forms of the energy-momentum tensors of
noncommutative field theories. The non-vanishing of the total
divergences of the energy-momentum tensors of noncommutative field
theories just reflect that energy-momentum are not locally conserved
on noncommutative spacetime. At the same time, from the
constructions of the quantum spacetime, we propose that the total
spacetime dimension of the quantum spacetime is infinite.

  However, over a long time, it is considered that
energy-momentum are locally conserved and energy-momentum
conservation is a fundamental principle of physics. But now, it is
possible that such a conclusion may be changed in noncommutative
field theories. In noncommutative spacetime, the law of
energy-momentum conservation may not satisfy. As we know that
noncommutative field theories have many different properties from
those of field theories on commutative spacetime, such as causality,
unitarity, and locality etc., we consider that energy-momentum
non-conservation is another property of noncommutative field
theories different from that of field theories on commutative
spacetime.

  From (55)--(58) for noncommutative scalar field theory and
noncommutative gauge field theory, we can see that the violation of
energy-momentum conservation is very weak because $\theta^{\mu\nu}$
is of order $l_{P}^{2}$ as postulated in the literature. Thus the
effect of energy-momentum non-conservation can only be obvious when
the observed spacetime scale is comparable to the Planck scale. The
energy-momentum non-conservation is a phenomenon of matter near the
Planck scale where the spacetime quantization effect cannot be
neglected. At the spacetime scale much larger than the Planck scale,
the noncommutativity of spacetime coordinates is not obvious, the
Moyal $\star$-product can be replaced by the ordinary product, and
the physics can be described by the field theories on commutative
spacetime approximately, thus energy-momentum are locally conserved
approximately. Or we can say that energy-momentum conservation is an
approximate phenomenon of physics in the spacetime scale much larger
than the Planck scale. On the other hand, under certain circumstance
and conditions, if the noncommutative parameters can be enlarged,
the effect of energy-momentum non-conservation can be enlarged. In
addition, we can see that in \cite{22}, from certain concrete models
of noncommutative field theories, the conclusion of energy-momentum
non-conservation has been derived in quantum scattering processes.
However, the purpose of this paper is to propose that
energy-momentum are non-conservative on noncommutative spacetime
from the expressions of energy-momentum tensors of noncommutative
field theories. The starting point of this paper is different from
those of \cite{22}.

  For the energy-momentum tensor problem of noncommutative field
theories, people have tried to search for some modified forms for
the energy-momentum tensors of noncommutative field theories.
However, it seems that the solutions are not satisfied.
Energy-momentum conservation is a conviction of physics. It seems
that such a conception is related with a recognition on the matter
of people, which is that the matter have the absolute objective
reality. If the matter have the absolute objective reality, it is
reasonable that energy-momentum should possess the property of
conservation. However, whether the matter have the absolute
objective reality really seems doubtful. This is because when the
spacetime is quantized, the classical meaning of the spacetime does
not exist any longer. Spacetime coordinates have become
noncommutative operators. Or we can say that the traditional concept
of the spacetime depending on the experience does not exist when the
spacetime is quantized. Thus the spacetime does not have the
absolute objective reality when it is quantized. It is difficult to
imagine that the matter whose existence relies on the existence of
spacetime can possess the absolute objective reality when the
spacetime itself does not possess the absolute objective reality.
Thus we consider that the matter whose existence relies on the
existence of spacetime do not have the absolute objective reality
when the spacetime is quantized. We consider that the conception of
energy-momentum conservation is related with the conception that the
matter are absolute objective reality. When the conception that the
matter are absolute objective reality is doubtful, the conception of
energy-momentum conservation will also be doubtful. Therefore we
propose that energy-momentum are non-conservative on noncommutative
spacetime; the energy-momentum tensors of noncommutative field
theories constructed from the Noether approach in subsection 4.3 are
just the correct forms for the energy-momentum tensors of
noncommutative field theories.

  In section 5, from the constructions of quantum spacetime and
spacetime quantization principle, we propose that the total
spacetime dimension is infinite, and it is quantized. Under the
meaning that the matter are existing and moving in an infinite
dimensional spacetime, the local conservation of energy-momentum can
be realized. For such an infinite dimensional spacetime, it seems
that there is no reason to assume that it is compactified for its
dimensions other than the first four. From the analysis of this
paper, we can see that there can at least exist the connections
between the matter in the observed four dimensional spacetime and
the additional higher dimensions through the exchange of the
energy-momentum between them.


\vskip 2cm

\begin{thebibliography}{99}

\baselineskip 12pt

\bibitem {1} H.S. Snyder, {\sl Quantized spacetime}, {\sl Phys.
Rev.} {\bf 71} (1947) 38.

\bibitem {2} S. Doplicher, K. Fredenhagen and J.E. Roberts,
{\sl Space-time quantization induced by classical gravity}, {\sl
Phys. Lett.} {\bf B 331} (1994) 39; {\sl The quantum structure of
space-time at the Planck scale and quantum fields}, {\sl Commun.
Math. Phys.} {\bf 172} (1995) 187 [hep-th/0303037].

\bibitem {3} M. Li and T. Yoneya, {\sl D-Particle dynamics and
the space-time uncertainty relation}, {\sl Phys. Rev. Lett.} {\bf
78} (1997) 1219 [hep-th/9611072].

\bibitem {4} N. Seiberg and E. Witten, {\sl String theory and
noncommutative geometry}, {\sl JHEP} {\bf 09} (1999) 032
[hep-th/9908142].

\bibitem {5} M.R. Douglas and N.A. Nekrasov, {\sl Noncommutative
field theory}, {\sl Rev. Mod. Phys.} {\bf 73} (2001) 977
[hep-th/0106048].

\bibitem {6} R.J. Szabo, {\sl Quantum field theory on
noncommutative spaces}, {\sl Phys. Rep.} {\bf 378} (2003) 207
[hep-th/0109162].

\bibitem {7} A. Gerhold, J. Grimstrup, H. Grosse, L. Popp,
M. Schweda and R. Wulkenhaar, {\sl The energy-momentum tensor on
noncommutative spaces - some pedagogical comments}, hep-th/0012112.

\bibitem {8} A. Micu and M.M. Sheikh-Jabbari, {\sl Noncommutative
$\Phi^4$ theory at two loops}, {\sl JHEP} {\bf 01} (2001) 025
[hep-th/0008057].

\bibitem {9} M. Abou-Zeid and H. Dorn, {\sl Comments on the
energy-momentum tensor in non-commutative field theories}, {\sl
Phys. Lett.} {\bf B 514} (2001) 183 [hep-th/0104244].

\bibitem {10} J.M. Grimstrup, B. Kloib\"{o}ck, L. Popp, V. Putz,
M. Schweda and M. Wickenhauser, {\sl The energy-momentum tensor in
noncommutative gauge field models}, {\sl Int. J. Mod. Phys.} {\bf A
19} (2004) 5615 [hep-th/0210288].

\bibitem {11} A. Das and J. Frenkel, {\sl On the energy-momentum
tensor in non-commutative gauge theories}, {\sl Phys. Rev.} {\bf D
67} (2003) 067701 [hep-th/0212122].

\bibitem {12} J.B. Geloun and M.N. Hounkonnou, {\sl Energy-momentum
tensors in renormalizable noncommutative scalar field theory}, {\sl
Phys. Lett.} {\bf B 653} (2007) 343.

\bibitem {13} H. Grosse and R. Wulkenhaar, {\sl Power-counting
theorem for non-local matrix models and renormalisation}, {\sl
Commun. Math. Phys.} {\bf 254} (2005) 91 [hep-th/0305066]; {\sl
Renormalisation of $\phi^{4}$-theory on noncommutative $R^{4}$ in
the matrix base}, {\sl Commun. Math. Phys.} {\bf 256} (2005) 305
[hep-th/0401128].

\bibitem {14} D.J. Gross, A. Hashimoto and N. Itzhaki, {\sl
Observables of non-commutative gauge theories}, {\sl Adv. Theor.
Math. Phys.} {\bf 4} (2000) 893 [hep-th/0008075].

\bibitem {15} Y. Okawa and H. Ooguri, {\sl Energy-momentum tensors
in matrix theory and in noncommutative gauge theories},
hep-th/0103124.

\bibitem {16} C.S. Chu, K. Furuta and T. Inami, {\sl Locality,
causality and noncommutative geometry}, {\sl Int. J. Mod. Phys.}
{\bf A 21} (2006) 67 [hep-th/0502012].

\bibitem {17} C.E. Carlson, C.D. Carone and N. Zobin, {\sl
Noncommutative gauge theory without Lorentz violation}, {\sl Phys.
Rev.} {\bf D 66} (2002) 075001 [hep-th/0206035].

\bibitem {18} R. Banerjee, B. Chakraborty and K. Kumar, {\sl
Noncommutative gauge theories and Lorentz symmetry}, {\sl Phys.
Rev.} {\bf D 70} (2004) 125004 [hep-th/0408197].

\bibitem {19} J.M. Romero, J.D. Vergara and J.A. Santiago,
{\sl Noncommutative spaces, the quantum of time and the Lorentz
symmetry}, {\sl Phys. Rev.} {\bf D 75} (2007) 065008
[hep-th/0702113].

\bibitem {20} A. Hashimoto and N. Itzhaki, {\sl Traveling faster
than the speed of light in non-commutative geometry}, {\sl Phys.
Rev.} {\bf D 63} (2001) 126004 [hep-th/0012093].

\bibitem {21} B. Durhuus and T. Jonsson, {\sl Noncommutative
waves have infinite propagation speed}, {\sl JHEP} {\bf 10} (2004)
050 [hep-th/0408190].

\bibitem {22} A.P. Balachandran, T.R. Govindarajan, A.G. Martins and
P. Teotonio-Sobrinho, {\sl Time-space noncommutativity: quantised
evolutions}, {\sl JHEP} {\bf 11} (2004) 068 [hep-th/0410067];

A.P. Balachandran, A.G. Martins and P. Teotonio-Sobrinho, {\sl
Discrete time evolution and energy nonconservation in noncommutative
physics}, {\sl JHEP} {\bf 05} (2007) 066 [hep-th/0702076].


\end{thebibliography}
\end{document}